\documentclass[final,3p,times]{elsarticle}

\usepackage{amsmath}
\usepackage{amssymb}
\usepackage{amsfonts}
\usepackage{xcolor}
\usepackage{adjustbox}
\usepackage{array}  % Fancier tables, controlling column width

\newcolumntype{L}[1]{>{\raggedright\let\newline\\\arraybackslash\hspace{0pt}}m{#1}}
\newcolumntype{C}[1]{>{\centering\let\newline\\\arraybackslash\hspace{0pt}}m{#1}}
\newcolumntype{R}[1]{>{\raggedleft\let\newline\\\arraybackslash\hspace{0pt}}m{#1}}

\journal{Nuclear Instruments and Methods in Physics Research B: Beam Interactions with Materials and Atoms}

\begin{document}

\begin{frontmatter}

%% Title, authors and addresses

%% use the tnoteref command within \title for footnotes;
%% use the tnotetext command for theassociated footnote;
%% use the fnref command within \author or \address for footnotes;
%% use the fntext command for theassociated footnote;
%% use the corref command within \author for corresponding author footnotes;
%% use the cortext command for theassociated footnote;
%% use the ead command for the email address,
%% and the form \ead[url] for the home page:
%% \title{Title\tnoteref{label1}}
%% \tnotetext[label1]{}
%% \author{Name\corref{cor1}\fnref{label2}}
%% \ead{email address}
%% \ead[url]{home page}
%% \fntext[label2]{}
%% \cortext[cor1]{}
%% \address{Address\fnref{label3}}
%% \fntext[label3]{}

\title{Smart energy models for atomistic simulations using a DFT-driven multifidelity approach}

%% use optional labels to link authors explicitly to addresses:
%% \author[label1,label2]{}
%% \address[label1]{}
%% \address[label2]{}

\author[kth,lugano]{Luca Messina}
\address[kth]{KTH Royal Institute of Technology, Nuclear Engineering, SE-106 91 Stockholm, Sweden}
\author[lugano]{Alessio Quaglino}
\address[lugano]{Institute of Computational Science, Universit\`a della Svizzera Italiana, CH-6900 Lugano, Switzerland}
\author[cea]{Alexandra Goryaeva}
\address[cea]{DEN -- Service de Recherches de M\'etallurgie Physique, CEA, Universit\'e Paris-Saclay, F-91191 Gif-sur-Yvette, France}
\author[cea]{Mihai-Cosmin Marinica}
\author[edf]{Christophe Domain}
\address[edf]{D\'epartement Mat\'eriaux et M\'ecanique des Composants, EDF–-R\&D, Les Renardi\`eres, F-77250 Moret-sur-Loing, France
}
\author[sck]{Nicolas Castin}
\address[sck]{Studie Centrum voor Kerneenergie, Centre d'\'Études de l'\'energie Nucl\'eaire (SCK\textbullet CEN), NMS Unit, B-2400, Mol, Belgium}
\author[sck]{Giovanni Bonny}
\author[lugano]{Rolf Krause}

\begin{abstract}

The reliability of atomistic simulations depends on the quality of the underlying energy models providing the source of physical information, for instance for the calculation of migration barriers in atomistic Kinetic Monte Carlo simulations. Accurate (high-fidelity) methods are often available, but since they are usually computationally expensive, they must be replaced by less accurate (low-fidelity) models that introduce some degrees of approximation. Machine-learning techniques such as artificial neural networks are usually employed to work around this limitation and extract the needed parameters from large databases of high-fidelity data, but the latter are often computationally expensive to produce. This work introduces an alternative method based on the multifidelity approach, where correlations between high-fidelity and low-fidelity outputs are exploited to make an educated guess of the high-fidelity outcome based only on quick low-fidelity estimations, hence without the need of running full expensive high-fidelity calculations. With respect to neural networks, this approach is expected to require less training data because of the lower amount of fitting parameters involved. The method is tested on the prediction of \textit{ab initio} formation and migration energies of vacancy diffusion in iron-copper alloys, and compared with the neural networks trained on the same database.

\end{abstract}

\begin{keyword}
%% keywords here, in the form: keyword \sep keyword
Machine learning \sep Multifidelity \sep Kinetic Monte Carlo \sep Atomistic simulations \sep Iron-Copper alloys
%% PACS codes here, in the form: \PACS code \sep code

%% MSC codes here, in the form: \MSC code \sep code
%% or \MSC[2008] code \sep code (2000 is the default)

\end{keyword}

\end{frontmatter}

%% \linenumbers

%% main text
\section{Introduction}

Numerical simulations play an important role in understanding and predicting the phenomena driving the microstructure evolution of solid materials. These processes cover such a wide range of length and time scales that a true multiscale approach is necessary. Typically, electronic-structure calculations based on Density Functional Theory (DFT) are used to parameterize atomic-scale simulations, which in turn open the way to higher scales thanks to various modeling techniques \cite{becquart:2010mm, knaster:2016a}. The reliability of these simulations crucially depends on the transfer of physical knowledge from the lower to the higher scale, which is often difficult to achieve given the intrinsic multifold complexity of materials at the electronic level. For this reason, several approximations are usually put in place. For instance, in atomistic kinetic Monte Carlo (AKMC) as well as in other methods it is necessary to compute the migration energy of given defects depending on the composition of the local atomic environment (LAE) \cite{becquart:2010kmc}. Nudged-elastic band (NEB) DFT calculations \cite{henkelman:2000a, henkelman:2000b} allow for their accurate calculation, but due to the high computational cost, it is certainly impossible to cover even a small fraction of the possible combinations. More generally, the same issue arises when a so-called \textit{energy model} is needed to predict the energy corresponding to a given atomic configuration, when the amount of possible configurations is so large that accurate calculations for all of them are out of reach.

This common parameterization problem is found in many branches of science: the parameters could be obtained with \textit{high-fidelity} models that are very accurate, but so computationally expensive that their use is impractical. Therefore, they are replaced by \textit{low-fidelity} models that introduce a certain degree of approximation but can provide all needed parameters with little computational effort \cite{peherstorfer:2016}. In atomistic simulations, DFT is the most typical high-fidelity choice, but for computational reasons it must be paired with low-fidelity models that make use of a limited set of DFT data. The most common examples in AKMC simulations are pair-interaction models \cite{soisson:2007, ngayamHappy:2013, huang:2017}, cluster-expansion developments \cite{lavrentiev:2010, wrobel:2017}, or interatomic potentials \cite{pareige:2011, bonny:2013}. Clearly, the main drawback is the loss of physical accuracy due to the involved approximations. These low-fidelity models are usually devised to correctly describe part of the physical properties, at the expense of others. Furthermore, it is very difficult to employ many of them at once: for instance, for the same alloy it is common to have several interatomic potentials tackling different properties, but choosing one of them entails discarding all the others.

In this context, machine learning (ML) algorithms can be very useful to extract data from the potentially large but for the most part inaccessible source of high-fidelity physical information, in a form that can be directly applied to the parameterization of higher-scale models \cite{curtarolo:2013}. It is indeed possible to create ML tools able to "learn" the physical properties from a limited dataset of high-fidelity examples (the \textit{training set}), and provide accurate estimations for unknown cases. So far, this has been achieved mainly with artificial neural networks (ANN), trained on DFT data and used to construct interatomic potentials for molecular dynamics (MD) \cite{artrith:2013, behler:2014} or KMC simulations \cite{castin:2017}, or for the direct prediction of migration energies \cite{castin:2011, castin:2012, messina:2017}. The latter works have proven that ANNs successfully achieve a full transfer of physical information from DFT to the higher scales, but they have also shown a few limitations. Firstly, many training examples are needed, so that using DFT as a source of training data is computationally demanding. In addition, the ANN is a sort of "black box" whose mathematical parameters are completely detached from the physical properties they are attempting to model. For this reason, the estimation of the target quantities can be very accurate, but there is limited room for understanding the physics behind it, and very little leveraging in case the simulation results are not correct. 

This work suggests an alternative method to ANNs to achieve the same goal, i.e., maximizing the use of high-fidelity data to ensure a full transfer of information across modeling scales. It relies on a \textit{multifidelity} (MF) \textit{approach} \cite{peherstorfer:2016}: instead of interpolating in a large database of high-fidelity data, this approach seeks a correlation between a limited dataset of high-fidelity calculations and the corresponding output of one or more low-fidelity models. If such a correlation is found, it is possible to provide "on-the-fly" estimations without the need of performing full and time-consuming high-fidelity calculations at each time. To the authors' knowledge, this method has never been applied to the parameterization of microstructure evolution simulations. With respect to ANNs, it is expected that comparable prediction accuracies can be reached with less training data points, with significant savings of computational time. In addition, low-fidelity models are combined together, rather than discarded, and each of them contributes to improving the prediction accuracy. Since different models can target different physical properties, MF models are expected to be more flexible and "physics-aware" than ANNs.

As a first test case, the MF approach is applied here to a DFT database of vacancy migration barriers in FeCu alloys, that was used to train the ANNs of previous works \cite{messina:2017, castin:2017}. The model is trained to predict the DFT energies and migration barriers of unknown atomic configurations and jump events, based on the estimation provided by some low-fidelity models \cite{pasianot:2007, soisson:2007}. If successful, this approach can lead to efficient DFT-aware energy models for KMC simulations, as well as for a wide range of methods requiring the estimation of formation or migration energies with little computational effort.

\section{Method}
\label{sec:method}

\subsection{Multifidelity framework}

Multifidelity methods are a class of techniques that accelerate calculations by leveraging on the correlations between the output of accurate but computationally expensive models, and that of inexpensive but less accurate ones \cite{peherstorfer:2016}. Mostly used in uncertainty quantification \cite{peherstorfer:2015}, they have been recently successfully applied to personalized medicine, in particular to cardiac electrophysiology \cite{quaglino:2018} and cardiovascular modeling \cite{biehler:2018}. This work aims at testing a similar technique on the parameterization of atomistic microstructure-evolution simulations. In what follows, a summary of the general mathematical framework is provided, and the reader is referred to \cite{peherstorfer:2016} for a more detailed description.

In the MF context, a \textit{model} is defined as a function $f :\mathcal{D} \rightarrow \mathcal{Y}$ that maps an input to an output, where $\mathcal{D} \subseteq \mathbb{R}^d$ is the input domain ($d \in \mathbb{N}$), and $\mathcal{Y} \subseteq \mathbb{R}$ the output domain. Let $z \in \mathcal{D}$ be the input and $y \in \mathcal{Y}$ the output. A \textit{high-fidelity} model, denoted with $f^{(1)}$, yields an accurate approximation
of the output of interest, whereas several \textit{low-fidelity} models, denoted with $f^{(k)}$ ($k \in \mathbb{N}_{>1}$), provide less accurate estimations of the same output. 
%We assume that the models $f^{(k)}$ are ordered with descending computational cost. 
Typically, since evaluating $y = f^{(1)}(z)$ is computationally expensive, it is unfeasible to sample $y_i$ at many inputs of interest $z_i$. Therefore, $f^{(1)}$ is usually replaced with $f^{(k)}$ for one suitable choice of $k$ that gives a sufficiently low error. Standard strategies to generate low-fidelity models are simplified models, projection methods, and data-fit surrogates \cite{peherstorfer:2016}. 

However, this approach presents three main limitations. First, the low-fidelity replacement is often unable to meet the desired accuracy. Secondly, when generating a low-fidelity model $f^{(k)}$ by means of projection or data fit with a training set $(z_i,f^{(1)} (z_i))_{i=1,\cdots,T}$, the amount of training data points needs to be large even for low-dimensional problems ($d \approx 15$). Finally, by selecting one low-fidelity model and discarding all the others, a large amount of information is lost. MF methods aim at overcoming such limitations by switching from model \emph{selection} to model \emph{fusion}, i.e., by combining all models to produce a better estimate of $y$. This can be done for instance via control variates \cite{peherstorfer:2015} or Bayesian regressions \cite{koutsourelakis:2009}. This study is focused on the latter approach, and in particular on its variant based on a Gaussian Process Regression (GPR) \cite{rasmussen:2006}.

The key idea is to improve the low-fidelity estimates by relying on the statistical dependency between $f^{(k)}$ and $f^{(1)}$, instead of minimizing the errors $\left| f^{(k)}(z)-f^{(1)}(z) \right|$. Mathematically, this can be achieved by creating an output-to-output data-fit surrogate:
\begin{equation} 
	y(z) = \mathcal{G}\left[f^{(2)}(z),\dots,f^{(m+1)}(z)\right] + \varepsilon, \quad \varepsilon \sim \mathcal{N}(0,\sigma) \; ,
	\label{eq:mf_fit}
\end{equation}
using a training set $[f^{(k)} (z_i),f^{(1)} (z_i)]$, with $i=1,...,T$ and $k=2,...,m+1$. In the GPR context, $\mathcal{G}$ is assumed to follow a GP distribution with a prescribed covariance kernel, whose parameters are fitted by maximizing a given likelihood. The underlying assumption is that the models $f^{(k)}$ contain partial information on the full model $f^{(1)}$, and that the missing information can be modeled as a normally-distributed systematic uncertainty term. Therefore, the availability of a large number of models contributes to better explain the total output variability, thus reducing the size of the training set. The accuracy of the fit can be then improved by including more low-fidelity models, as an alternative to adding more training data points. Moreover, the dimensionality and complexity of the model in Eq. (\ref{eq:mf_fit}) are typically lower than those of the input. Hence, a small training set is sufficient to achieve an accurate data fit, as opposed to the case of input-to-output surrogates such as ANNs, which are in general more involved. Furthermore, the use of a Bayesian regression complements the estimate with confidence intervals, which can provide error indicators for refining the low-fidelity models, if necessary.

\subsection{Application to vacancy migration in FeCu}

The multifidelity approach is tested on the modeling of single-vacancy diffusion in FeCu dilute alloys, applicable for instance to the parameterization of AKMC simulations. The high-fidelity model is a large DFT database previously produced to train an ANN \cite{messina:2017}, as well as to create fully ANN-based interatomic potentials \cite{castin:2017}. The database consists of approximately 2000 NEB calculations \cite{henkelman:2000a, henkelman:2000b} of vacancy migration in 249-atom supercells, half of them featuring a perfect solid solution with a random distribution of Cu atoms (up to 5 at.\%), and the other half snapshots from previous AKMC simulations \cite{pasianot:2007, messina:2017} with small Cu clusters. The 2000 cases were selected in order to obtain the most diverse database in terms of jumping atom type (Fe or Cu), local atomic environment (LAE) around the vacancy, and migration barrier values. The calculations were performed with the Vienna
\textit{ab initio} simulation package ({\sc vasp}) \cite{kresse:1993, kresse:1994a, kresse:1994b} using the projector-augmented wave (PAW) pseudopotentials \cite{blochl:1994, kresse:1999} in the Perdew-Burke-Erzernhof (PBE) approximation \cite{perdew:1996}. Further details on the \textit{ab initio} calculations can be found in \cite{messina:2017}.  

The low-fidelity inputs are given by two independent energy models used in the past to investigate diffusion and precipitation of Cu in Fe alloys: an interatomic potential based on the embedded-atom method (EAM) \cite{pasianot:2007}, and a broken-bond pair-interaction model by Soisson \textit{et al.} \cite{soisson:2007}. The latter yields the migration barrier of a given migration event as a function of the LAE of the vacancy and the jumping atom, limitedly to first- and second-nearest neighbors. On the other hand, the EAM potential is used to compute the relaxed supercell energy in the initial and final configuration for each of the 2000 migration events, and to perform the corresponding NEB calculations  following the same strategy and parameterization of previous works \cite{castin:2011, castin:2012}.

The multifidelity fitting is aimed at estimating:
\begin{enumerate}
	\item The supercell formation energy $E^\mathrm{f}$, computed from the supercell total energy $E^*$ as:
	\begin{equation}
	E^\mathrm{f} = E^* - \left[ E^\mathrm{c}_\mathrm{Fe} N_\mathrm{Fe} + E^\mathrm{c}_\mathrm{Cu} N_\mathrm{Cu}  \right] \; , 
	\label{eq:form_energy}
	\end{equation}
	where $E^\mathrm{c}_\mathrm{Fe}$ and $E^\mathrm{c}_\mathrm{Cu}$ are the cohesive energy per atom in pure bcc Fe and pure bcc Cu obtained with either DFT ($-8.31$ and $-3.68$ eV) or EAM ($-4.12$ and $-3.49$ eV).
\item The migration barriers $E^\mathrm{mig}_{\mathrm{i}\rightarrow \mathrm{f}}$ and $E^\mathrm{mig}_{\mathrm{f}\rightarrow \mathrm{i}}$, taken as the difference between the supercell energy at the saddle point and that in the initial or final state, respectively.
	\item The energy difference between the final (f) and initial (i) state: $\Delta E = E_\mathrm{f} - E_\mathrm{i}$.
	\item The saddle-point energy computed from the migration barrier as:
	\begin{equation}
	E^\mathrm{sad} = E^\mathrm{mig}_{\mathrm{i}\rightarrow \mathrm{f}} - \frac{\Delta E}{2}  \; .
	\label{eq:mig_ene_rescaling}
	\end{equation}
\end{enumerate}
The rescaling in Eq. (\ref{eq:form_energy}) is an arbitrary choice aimed at reducing the variability of the fitting target. Reformulating the migration barrier as in Eq. (\ref{eq:mig_ene_rescaling}) allows for a clear distinction between the thermodynamic and kinetic contribution, and the consequent possibility of targeting the two aspects with different high-fidelity models, if needed \cite{messina:2017, castin:2017}. In this specific application, such splitting is advisable because the PAW-PBE functionals underestimate the solubility of Cu in Fe due to a faulty prediction of the solution energy \cite{olsson:2010}. Note that with the definition in Eq. (\ref{eq:mig_ene_rescaling}), the saddle energy is identical for the forward and backward jumps.

For each of the target quantities ($E^\mathrm{f}$, $E^\mathrm{mig}$, $\Delta E$, $E^\mathrm{sad}$), a Gaussian regression of the type in Eq. (\ref{eq:mf_fit}) is performed, with varying sets of low-fidelity models $f^{(k)}(z)$. The data are regressed with a multidimensional radial-basis function (RBF) kernel and a noise function \cite{rasmussen:2006}. 

\subsection{Descriptors}

In most MF applications, the low-fidelity outputs are directly correlated with the high-fidelity ones. However, the energy associated to an arrangement of atoms depends in a highly non-linear fashion on the position of each atom. The use of descriptors is therefore essential to include information about atomic positions in the fitting. A brief explanation is provided here, and the reader is referred to a more general description of the descriptor framework in \cite{bartok:2013}.

Descriptors are symmetry-invariant mathematical representations of a crystal structure that replace the conventional Cartesian coordinates while leaving the physical properties unchanged. They allow the essential features of the LAE to be comprised into a space of lower dimensionality, known as the \textit{descriptor space}. In this framework, each atom in the supercell is assigned a series of coefficients (one for each descriptor) based on the surrounding LAE. These coefficients establish a correlation between the atomic configuration and the corresponding physical properties (e.g., the supercell energy), allowing for some kind of regression to be performed.

This study relies exclusively on Gaussian regressions of the type in Eq. (\ref{eq:mf_fit}), and features the use of the spectral descriptor SO(4) bispectrum \cite{bartok:2013, bartok:2009}. The latter was shown to be complete \cite{kondor:2007, kakarala:1992}, i.e., able to describe LAEs uniquely, including all local symmetry operations such as translation, rotation, and reflection. Instead of using the atomic coordinates $\mathbf{r}$, the LAE around the $a^\mathrm{th}$ atom is described with a neighbor density function  $\rho_a(\textbf{r})$, corresponding to the bispectrum components of the four-dimensional (4--D) hyperspherical harmonics projected onto the $\mathcal{R}^3$-sphere ($\theta_{0}$, $\theta$, $\phi$) \cite{varshalovich:1988}. The angular components are projected onto a spherical harmonic function defined by the polar angles $\theta$ and $\phi$, and the radial component is converted into the third polar angle $\theta_{0}$. The relation between polar and Cartesian coordinates is bijective.
%It is interesting to note that these functions are the transformations matrices for standard spherical harmonics under rotations by angle $\varphi = 2 \theta_0$ around the axis defined by the angles $\theta$ and $\phi$.
%Consequently, these functions can be easily related to the Wigner $D$-matrix, $D^j_{m, m'}$ (which gives the irreducible representations of the SO(3) group \cite{angmom}).

The neighbor density function is expanded in 4-D hyperspherical harmonics $U_{jmm'}$ as follows:
\begin{equation}
\label{eq:density}
\rho_a(\textbf{r}) = \sum_{b \, \in \, \mathcal{V}(a) } w_b \, \delta \, (\mathbf{r} - \mathbf{r}_{ab}) = \sum_{j=0}^{\infty} \sum_{m=-j}^j \sum_{m'=-j}^j c_{jm m'}^a
U_{jmm'} (\theta_0, \theta, \phi), %\; \; j = 0,\frac{1}{2},1,\frac{3}{2}, ... \; ,
\end{equation}
where the sum runs over all neighbors $b\in \mathcal{V}(a)$ of the $a^\mathrm{th}$ atom within a cutoff distance $R_\mathrm{cut}$, and $w_b$ is an arbitrary weight associated with the chemical species of $b$ (in this work: $R_\mathrm{cut} = 5 $ a$_0$, $w_\mathrm{Fe} = 1$, and $w_\mathrm{Cu} = 2$). The power spectrum coefficients $c_{jm m'}^a$ are computed as the inner product between the density and the hyperspherical functions. Index $j$ can take only positive integer or half-integer values ($ j = 0,\frac{1}{2},1,\frac{3}{2}, ...$), and is limited for practical purposes to a maximal value $j_\mathrm{max}=7/2$. The latter choice guarantees a satisfactory compromise between numerical precision and computational load. %The total number of spherical harmonics is thus $(2j_\mathrm{max}+1)^3 = 512$. 

Once the $c_{j;m'm}^a$ coefficients are known, the SO(4) bispectrum coefficients $B_{ll_1l_2}^a$ \cite{bartok:2013} associated to each atom $a$ can be written as:

\begin{equation}
\label{eq:desc_bispectrum_so4}
B_{ll_1l_2}^a = \sum_{m',m=-l}^{l}\sum_{m_1',m_1=-l_1}^{l_1} \sum_{m_2',m_2=-l_2}^{l_2}
c_{j;m'm}^{a*}C_{m'm_1'm_2'}^{ll_1l_2}C_{mm_1m_2}^{ll_1l_2}c_{l_1; m_1'm_1}^{a}c_{l_2; m_2'm_2}^{a}
\end{equation}

where $C_{mm_1m_2}^{ll_1l_2}$ are the Clebsch--Gordan (CG) coefficients. From the total set of 512 coefficients for $j_\mathrm{max}=7/2$, the CG selection rules narrow them down to 30 coefficients only \cite{bartok:2009, thompson:2015}. However, Kakarala \cite{kakarala:1992} has shown that this choice is still overcomplete, and the coefficient set can be further restricted to the diagonal components ($l_1=l_2$) only, yielding 26 independent coefficients ($B^a_1, B^a_2, \dots, B^a_{26}$) for each atom \cite{thompson:2015, thompson:2017, thompson:2018}. This strategy, originally suggested by Varshalovich \textit{et al.} \cite{varshalovich:1988}, is implemented in the MiLaDy package \cite{milady:2018} used here to compute the descriptor coefficients. 

Finally, for each component $i$ the corresponding coefficients of all atoms are summed to obtain a global coefficient $D_i = \sum_{a} B^a_i$ (with $i=1,\dots,26$). The latter is used to seek a correlation with the supercell energy, % assuming a linear relation between energy and descriptor coefficients \cite{thompson:2015}. 
and can be thus regarded as a low-fidelity energy estimation that contains information about mutual atomic interactions. %although the actual number has nothing to do with an energy (and can in fact have a very different order of magnitude, depending on the component). 
This procedure hence yields 26 low-fidelity guesses that are added to the low-fidelity models (EAM and Soisson) mentioned in the previous section. For the targeted quantities obtained as energy differences or sums ($E^\mathrm{mig}$, $\Delta E$, $E^\mathrm{sad}$), the same transformations are applied to each $D_i$. For instance, the high-fidelity $\Delta E$ is correlated to $D_i^\mathrm{f} - D_i^\mathrm{i}$, and by analogy for the other quantities. This is to avoid uncertainty propagation that would occur when computing $\Delta E$ from the formation energies.

Building additional low-fidelity models with descriptors requires the availability of some low-fidelity estimations of the atomic positions. Interatomic potentials such as the EAM model do provide this information, as they allow for relaxation calculations, whereas rigid-lattice schemes such as pair-interaction models do not. However, positions corresponding to the rigid lattice can be taken as a rough low-fidelity guess of the atomic coordinates. This is tested in Fit E, among the fitting sessions listed in Table \ref{tab:fitting_sessions}.

\subsection{Fitting sessions}

Several fitting sessions are performed, first with the energy guesses from the EAM potential and the Soisson model ($E_\mathrm{EAM}$, $E_\mathrm{S}$), and then including the descriptor coefficients computed with the rigid-lattice or the EAM-relaxed atomic coordinates, respectively $D_{i=1,...,n}(x_\mathrm{rigid})$ and $D_{i=1,...,n}(x_\mathrm{EAM})$. The sessions are summarized in Table \ref{tab:fitting_sessions}. Sessions A and B are aimed at verifying weather the descriptor coefficients are effectively correlated with the supercell energy. The training data points are picked in a complete random fashion among the 2000 NEB cases,  or twice as many for $E^\mathrm{f}$ and $E^\mathrm{mig}$, while the remaining ones are used for validation. 

\begin{table*}[htb]\scriptsize
\caption{Validation results of the fitting sessions with Gaussian regression performed in this work, with a training set of 300 data points, and different low-fidelity sources. $E_\mathrm{EAM}$ and $E_\mathrm{S}$ denote the energy computed with the EAM potential \cite{pasianot:2007} or the Soisson model \cite{soisson:2007}, respectively, while $D_i(x)$ marks the 26 descriptor coefficients based on either the rigid positions ($x_\mathrm{rigid}$) or the EAM-relaxed positions ($x_\mathrm{EAM}$). $\Delta \varepsilon$ is the mean validation error. Session D was not performed on $E^\mathrm{f}$ because the Soisson model does not predict the formation energy. } 
\renewcommand{\arraystretch}{1.3}
\centering
\begin{adjustbox}{max width=\textwidth}
\begin{tabular}{C{0.5cm}C{1.3cm}C{3.3cm}|C{1cm}C{1cm}|C{1cm}C{1cm}|C{1cm}C{1cm}|C{1cm}C{1cm}}
	\hline  
	& & & & & & & & & & \\[-0.8em]
	Fit					& High-fidelity		& Low-fidelity 	&	\multicolumn{2}{c|}{Formation energy $\left( E^\mathrm{f} \right)$} & \multicolumn{2}{c|}{Migration barriers $\left( E^\mathrm{mig} \right)$}  &  \multicolumn{2}{c|}{Energy difference $\left( \Delta E \right)$}  &  \multicolumn{2}{c}{Saddle energy $\left( E^\mathrm{sad} \right)$}  \\
				&						&						&	$\Delta \varepsilon$ [meV] &	$R^2$	&	$\Delta \varepsilon$ [meV] &	$R^2$  &	$\Delta \varepsilon$ [meV] &	$R^2$  &	$\Delta \varepsilon$ [meV] &	$R^2$  \\
	\hline
		& & & & & & & & & & \\[-0.8em]
	\multicolumn{3}{l|}{\textbf{Descriptor performance}}  &   &  &  &  &  &  &  &  \\
	A                  & $E_\mathrm{EAM}$  & $D_i(x_\mathrm{EAM})$	 & $ 57.6 $ & $ 0.9980 $ & $ 34.3 $ &$ 0.9312 $ & $ 48.1 $ & $ 0.8951 $ &$ 19.1 $  & $ 0.9704 $  \\
	B                  & $E_\mathrm{DFT}$ & $D_i(x_\mathrm{DFT})$  & $ 194.8 $  & $ 0.9951$ & $ 67.4 $ & $ 0.8902$ & $ 113.8 $ & $0.9031 $ & $ 43.4 $ &  $ 0.7066$  \\
	\hline
	& & & & & & & & & & \\[-0.8em]
	\multicolumn{3}{l|}{\textbf{Multifidelity fitting}}  &   &  &  &  &  &  &  &  \\
	C                  & $E_\mathrm{DFT}$ & $E_\mathrm{EAM}$  	&  $ 941.6 $ & $ 0.8675 $ & $ 97.3 $ &$ 0.5610 $ & $ 102.7 $ & $ 0.8231$ &$ 62.2 $  & $ 0.4416 $   \\
	D                  & $E_\mathrm{DFT}$  & $E_\mathrm{EAM}$, $E_\mathrm{S}$  & --  & -- & $ 72.5 $ & $ 0.7624 $ & $ 74.1 $ & $ 0.9065 $ & $ 53.5 $ &  $ 0.5864 $	 \\
	E                  & $E_\mathrm{DFT}$ & $D_i(x_\mathrm{rigid})$  & $ 146.5 $  & $ 0.9972 $ & $ 56.0 $ & $ 0.9249 $ & $ 111.1 $ & $ 0.9087 $ & $ 42.0 $ & $ 0.7242 $ 	 \\
	F                  & $E_\mathrm{DFT}$ & $D_i(x_\mathrm{EAM})$  & $ 132.3 $  & $ 0.9974 $ & $ 59.6 $ &$ 0.8387 $  & $ 62.0 $ & $ 0.9348 $ &$ 40.6 $  & $ 0.7636 $ 	 \\
	G                  &  $E_\mathrm{DFT}$  & $E_\mathrm{EAM}$, $E_\mathrm{S}$, $D_i(x_\mathrm{rigid})$, $D_i(x_\mathrm{EAM})$ & $ 110.4 $  & $ 0.9982 $ & $ 51.6 $ & $ 0.8777 $ & $ 58.0 $ & $ 0.9433 $ & $ 37.2 $ &  $ 0.8008 $ 	 \\ 
	\hline   
\end{tabular}
\end{adjustbox}
\label{tab:fitting_sessions}
\end{table*}

\section{Results}

\subsection{Descriptor performance}

\begin{figure}[htb]
	\centering{}
	\includegraphics[width=0.6\columnwidth]{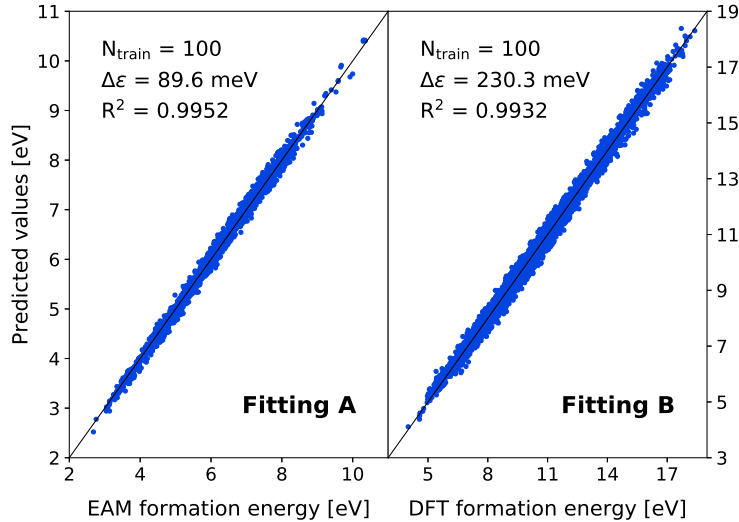}
	\caption{Validation results of the fitting session A (B), where the descriptor coefficients computed with the EAM-relaxed (DFT-relaxed) atomic positions are used to predict the EAM (DFT) supercell formation energy, based on a training set of 100 data points. $\Delta \varepsilon$ is the mean validation error.}
	\label{fig:descriptor_performance}
\end{figure}

If the descriptor coefficients correlate well with the target quantities, and in particular with the supercell formation energy, they can provide valuable additional low-fidelity estimations. This correlation is checked by performing a multi-dimensional Gaussian regression on the 26 descriptor coefficients computed on the relaxed atomic positions, using a varying amount of training data points ($N_\mathrm{train}$). The accuracy of the fitting is then validated on the remaining data. The validation results are shown in Table \ref{tab:fitting_sessions} for all target quantities with $N_\mathrm{train}=300$, and in Fig. \ref{fig:descriptor_performance} specifically for the supercell formation energy with $N_\mathrm{train}=100$. Fit A refers to the EAM energy and relaxed positions, while Fit B to the corresponding DFT quantities. 

Figure \ref{fig:descriptor_performance} shows that the predicted formation energies correlate well with the validation data in both fitting sessions, and this correlation is visible already for low $N_\mathrm{train}$. A weaker correlation is found for the other target quantities in Table \ref{tab:fitting_sessions}. This confirms that the descriptor coefficients can be used as source of low-fidelity data, which improves the MF fitting shown in the next section. However, the mean validation errors are higher than those obtained with ANNs. Regarding the DFT data, the ANN-based rigid-lattice potential \cite{castin:2017} trained on 1300 data points reached an accuracy on the formation energy of 0.191 meV/atom, or 48 meV per supercell, while in Fit B the mean error with $N_\mathrm{train}=1300$ (not shown) is 156 meV, and just slightly higher (195 meV) with $N_\mathrm{train}=300$. Therefore, if they were to replace computationally expensive DFT energy calculations, descriptors would need a limited amount of training data, but in this specific case the error ($\approx 0.2$ eV) might be too large for practical applications. The prediction of migration barriers is more accurate (55 meV with $N_\mathrm{train}=1300$, and 67 meV with $N_\mathrm{train}=300$), but still larger than the ANN performance (23.5 meV) \cite{messina:2017}. In Fit A, the accuracy is considerably better thanks to the intrinsic simplicity of the EAM model with respect to DFT, so that it is easier for descriptors to predict its outcome.  

%Since we have established that descriptors can produce several low-fidelity models to improve the multifidelity fitting, we will verify in the next section how that worked out. It is worth reminding that the objective is to use low-fidelity models to replace expensive high-fidelity calculations, and to this extent it is thus not possible to use the descriptor coefficients obtained with the DFT-relaxed atomic positions which would require a full DFT calculation to be obtained. Therefore, $D_i(x_\mathrm{DFT})$ cannot be used as low-fidelity input.

\subsection{Multifidelity fitting}

\begin{figure*}[htb]
	\centering{}
	\includegraphics[width=1\textwidth]{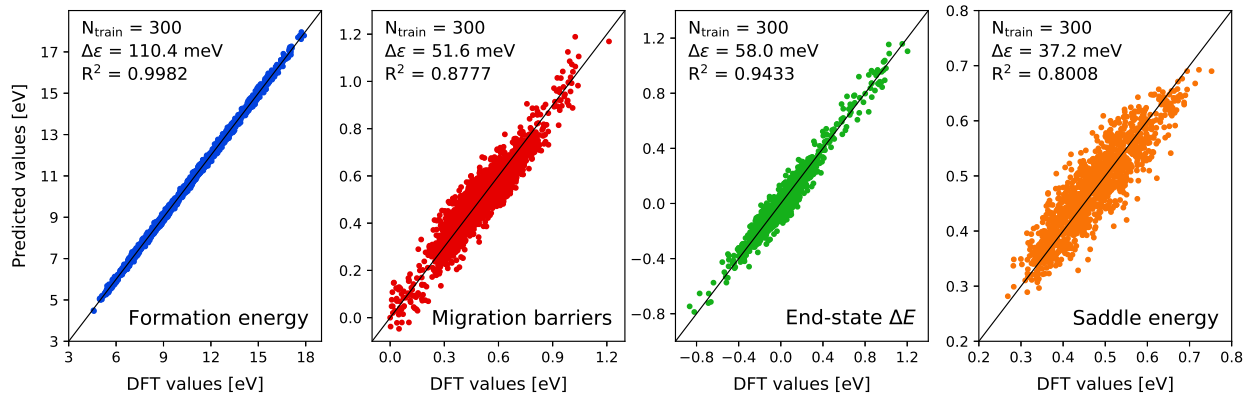}
	\caption{Validation results of the fitting session G, based on a training set of 300 data points, to predict the DFT target quantities (formation energy, migration barriers, energy difference, and saddle energy) with all available low-fidelity models: the EAM energy \cite{pasianot:2007}, the Soisson energy \cite{soisson:2007}, and the descriptor coefficients obtained with the rigid-lattice and the EAM-relaxed atomic positions. $\Delta \varepsilon$ is the mean validation error.}
	\label{fig:best_fit}
\end{figure*}

\begin{figure}[htb]
	\centering{}
	\includegraphics[width=0.6\columnwidth]{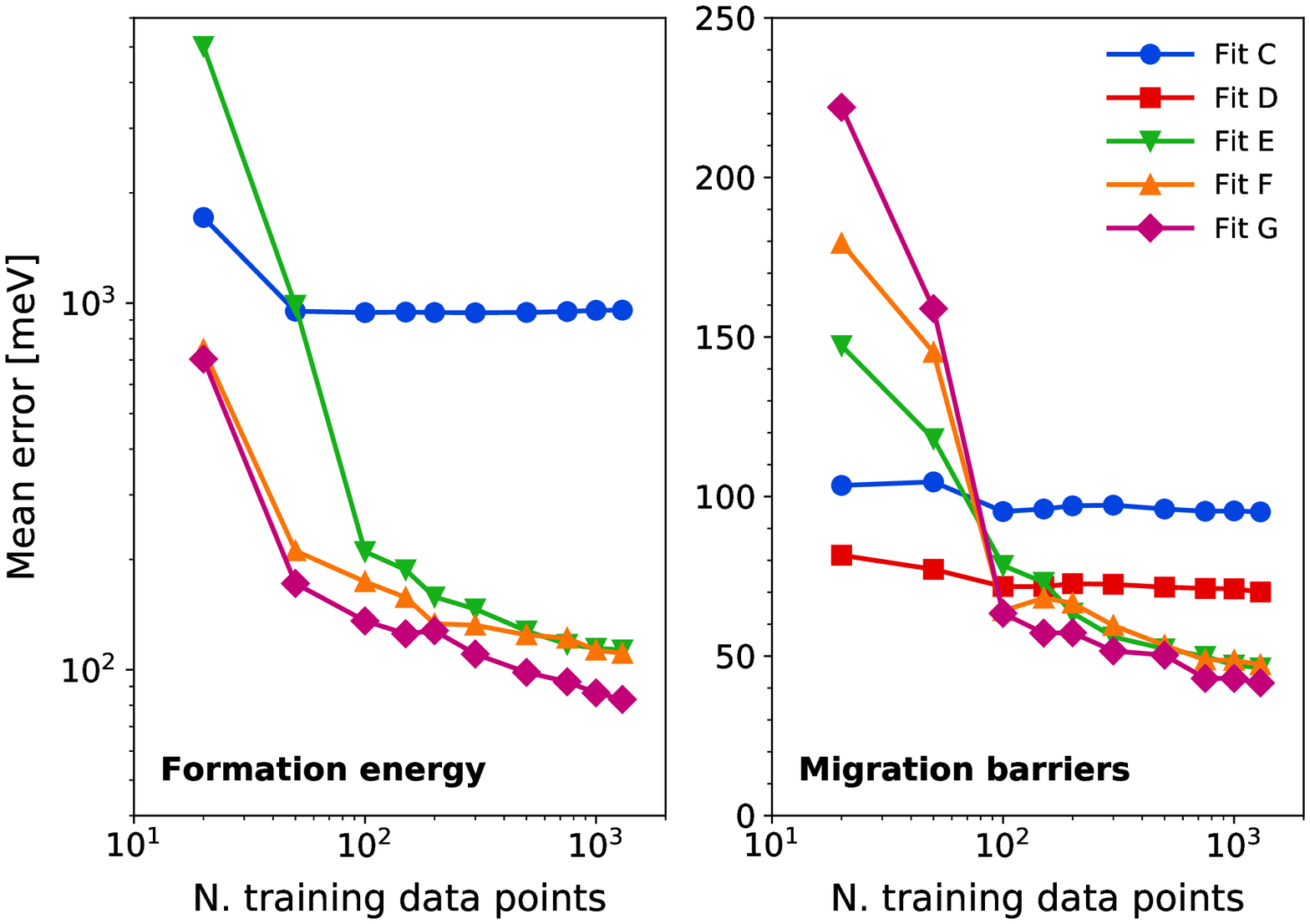}
	\caption{Mean validation error $\Delta \varepsilon$ as a function of the number of training data points, for the prediction of the formation energy (left) and the migration barriers (right).}
	\label{fig:varying_n_training}
\end{figure}

The MF framework described in Section \ref{sec:method} is applied to the prediction of the DFT target quantities with an increasing number of low-fidelity models: first with energy guesses only (Fit C and D), and then with the descriptor coefficients based on rigid-lattice positions (Fit E) and the EAM-relaxed positions (Fit F). Finally, all low-fidelity inputs are used in the last fitting (Fit G). Table \ref{tab:fitting_sessions} shows the validation results with $N_\mathrm{train}=300$, while Fig. \ref{fig:best_fit} visually depicts those obtained in Fit G. The variation of the mean validation error with increasing number of training data points is shown in Fig. \ref{fig:varying_n_training}. 

Without descriptors, the performance of the MF model trained only on energy estimations (Fit C and D) is very poor, with a $\approx 1$ eV mean error on the formation energy and low R-squared values on the other quantities. The addition of the Soisson model, which targets saddle-point interactions and is therefore expected to be more accurate on migration energies than the EAM model, improves the correlation and reduces the mean error, proving that the MF approach can successfully take advantage of several models at once. However, Fig. \ref{fig:varying_n_training} shows that increasing $N_\mathrm{train}$ does not lead to any gain of accuracy, and the error on the migration barriers is never lower than 0.07 eV even with $N_\mathrm{train}=1300$. The results without descriptors are thus fully unsatisfactory, and cannot be improved by adding more training examples, because the model is missing any information about how the atomic positions influence the formation and migration energies.

Including the descriptor coefficients in the low-fidelity input improves drastically the formation energy prediction, and moderately the migration barriers and the other quantities, with much better R-squared values. This confirms once more the strength of the MF approach, where the contributions of several low-fidelity models targeting different properties can be joined together to provide more accurate predictions. Fig. \ref{fig:varying_n_training} shows indeed that the accuracy can be increased by either increasing the amount of high-fidelity training data, as was the case for ANNs, or by including or producing more low-fidelity models.
%, possibly targeting different sides of the physical description of the system. In our case, we have an energy model (EAM) that was developed to reproduce well the TD properties, a model (Soisson) focused on the saddle-point kinetics, and a set of descriptor coefficients trying to catch the contribution of interatomic interactions to the total energy.
In addition, Fig. \ref{fig:varying_n_training} shows that a discrete level of accuracy is already reached with 200-300 training data points, and increasing the training set does not yield substantial improvements. This means that potentially, if a sufficient amount of low-fidelity models is available, the computational cost to produce high-fidelity data can be considerably reduced to approximately 1/4 or even 1/6 of the amount required for ANNs. 

\section{Discussion}

The results of this work prove that the MF framework can be successfully applied to build energy models for atomistic simulations, allowing for a higher degree of flexibility (by choosing or developing the needed low-fidelity models), and a likely substantial reduction of the computational requirements with respect to ANNs. If compared to the ANN accuracy with the same amount of training data ($N_\mathrm{train}=1300$), the MF accuracy is not as good: (83 vs 48 meV on the formation energy, and 41.6 vs 23.5 meV on the migration barriers) . However, the MF predictions are already quite accurate with much fewer data points ($N_\mathrm{train}=300$). The ANN performance with so few data points has not been tested in the corresponding previous works \cite{messina:2017, castin:2017}, but is likely to be poorer because of the high number of ANN parameters to be fitted. Since the MF accuracy could be improved by adding more low fidelity models, it is expected that for systems where several interatomic potentials exist the MF performance should be closer to the ANN one. 

Thanks to the general character of the MF approach, the method is certainly suitable for developing lattice-free potentials, for instance by including forces among the target quantities. It is also worth noting that in the specific case of FeCu alloys, the thermodynamic properties stemming out of the EAM potential should be more accurate than the DFT PAW-PBE functionals, but thanks to the splitting of the migration energy according to Eq. (\ref{eq:mig_ene_rescaling}), it could be possible to use $\Delta E_\mathrm{EAM}$ as a high-fidelity reference for the thermodynamic part, and $E^\mathrm{sad}_\mathrm{DFT}$ for the kinetic part. Finally, it should be mentioned that although faster than DFT calculations, computing the descriptor coefficients for a given atomic configuration still requires more time than a EAM calculation (in the order of a few seconds as opposed to milliseconds \cite{thompson:2017}). Therefore, the applicability of the method to KMC simulations with an "on-the-fly" low-fidelity engine featuring descriptors is still to be demonstrated.

\section{Conclusions}

This work presented multifidelity (MF) methods as an innovative machine-learning based approach for constructing energy models that can predict the static energy of a given arrangement of atoms, as well as the migration barriers of given jump events, to provide the relevant parameterization to atomistic simulations or analytical models while ensuring the best possible transfer of physical properties from \textit{ab initio} calculations. MF makes use of several approximate (low-fidelity) models to learn predicting the outcome of an accurate (high-fidelity) model without the need for performing the actual time-consuming calculation. The method, tested on the prediction of DFT formation energies and migration barriers in dilute FeCu supercells, proved to be successful in describing the correlation between the high-fidelity and the low-fidelity values. Although the performance in terms of prediction accuracy is not as good as artificial neural networks (ANN), the MF mean prediction errors are already reasonably low with much smaller training datasets than ANNs, which shows that potentially the amount of high-fidelity data needed for training can be considerably reduced, with important savings of computational power. Furthermore, it has been shown that the more low-fidelity models, the more accurate the results. MF methods are indeed capable of combining together the information given by several models, which ensures a high degree of flexibility, while keeping the needed amount of high-fidelity calculations to its minimum. In this context, the use of descriptors has been necessary to describe the influence of the local atomic environment on the energetics. In conclusion, this work has demonstrated the applicability of MF methods to multiscale materials modeling and microstructure evolution simulations. They can contribute to improving the reliability of atomistic simulations by maximizing the amount of accurate DFT properties that can be transferred, with little approximation, to the higher modeling scales.

\section*{Acknowledgments}
This work has been carried out within the framework of the EUROfusion Consortium and has received funding from the Euratom research and training programme 2014-2018 under grant agreement No 633053. A. Goryaeva acknowledges the financial support by the Cross-Disciplinary Program on Numerical Simulations of CEA, the French Alternative Energies and Atomic Energy Commission. The high-performance computational resources have been provided by \'Electricit\'e de France (R\&D).

%% The Appendices part is started with the command \appendix;
%% appendix sections are then done as normal sections
%% \appendix

%% \section{}
%% \label{}

%% If you have bibdatabase file and want bibtex to generate the
%% bibitems, please use
%%
%\bibliographystyle{elsarticle-num} 
%\bibliography{paper17_arXiv.bib}

%% else use the following coding to input the bibitems directly in the
%% TeX file.

\end{document}